\documentclass[10pt,journal,compsoc]{IEEEtran}

\ifCLASSOPTIONcompsoc
    \usepackage[nocompress]{cite}
\else
    \usepackage{cite}
\fi

\usepackage{graphicx}
\usepackage{cite}
\usepackage{xcolor}
\usepackage{url}

\usepackage[caption=false]{subfig}
\usepackage{fontawesome}
\usepackage{lipsum}
\usepackage{comment}
\usepackage{tabularx}
\usepackage{booktabs}
\usepackage{multirow}
\usepackage[normalem]{ulem}
\usepackage{ragged2e}

\newcommand{\edit}[1]{\textcolor{black}{#1}}
\newcommand{\remove}[1]{\textcolor{brown}{\sout{}}}

\newcommand{\speechicon}{\raisebox{-.1em}{\includegraphics[height=1em]{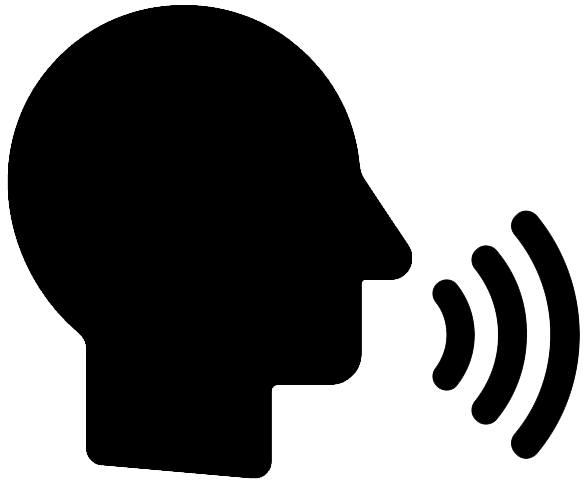}}}
\newcommand{\touchicon}{\raisebox{-.2em}{\includegraphics[height=1.15em]{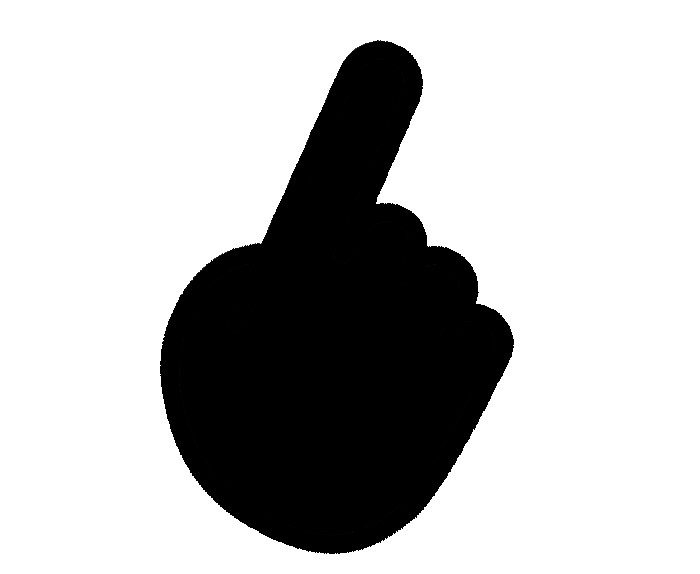}}}

\begin{document}

\title{Touch? Speech? or Touch and Speech? Investigating Multimodal Interaction \\for Visual Network Exploration and Analysis}

\author{Ayshwarya Saktheeswaran, Arjun Srinivasan, and John Stasko
\IEEEcompsocitemizethanks{\IEEEcompsocthanksitem Ayshwarya Saktheeswaran, Arjun Srinivasan, and John Stasko are with Georgia Institute of Technology, Atlanta,
GA, 30332.\protect\\
E-mail: \{ayshwarya6, arjun010\}@gatech.edu, stasko@cc.gatech.edu
}% <-this % stops an unwanted space
\thanks{Manuscript received December 20, 2019; revised January 23, 2020.}
}

\markboth{Journal of \LaTeX\ Class Files,~Vol.~14, No.~8, August~2015}%
{Shell \MakeLowercase{\textit{et al.}}: Bare Demo of IEEEtran.cls for Computer Society Journals}

\IEEEtitleabstractindextext{%
\begin{abstract}
Interaction plays a vital role during visual network exploration as users need to engage with both elements in the view (e.g., nodes, links) and interface controls (e.g., sliders, dropdown menus). Particularly as the size and complexity of a network grow, interactive displays supporting multimodal input (e.g., touch, speech, pen, gaze) exhibit the potential to facilitate fluid interaction during visual network exploration and analysis. While multimodal interaction with network visualization seems like a promising idea, many open questions remain. For instance, do users actually prefer multimodal input over unimodal input, and if so, why? Does it enable them to interact more naturally, or does having multiple modes of input confuse users? To answer such questions, we conducted a qualitative user study in the context of a network visualization tool, comparing speech- and touch-based unimodal interfaces to a multimodal interface combining the two. Our results confirm that participants strongly prefer multimodal input over unimodal input attributing their preference to: 1) the freedom of expression, 2) the complementary nature of speech and touch, and 3) integrated interactions afforded by the combination of the two modalities. We also describe the interaction patterns participants employed to perform common network visualization operations and highlight themes for future multimodal network visualization systems to consider.
\end{abstract}

}
\maketitle

% \IEEEdisplaynontitleabstractindextext
% \IEEEpeerreviewmaketitle

\IEEEraisesectionheading{\section{Introduction}\label{sec:introduction}}

\IEEEPARstart{N}{}etwork visualizations, often in the form of node-link diagrams, are useful for describing and exploring data relationships in many domains such as biology~\cite{mason2007graph}, the social sciences~\cite{moody2004structure}, and transportation planning~\cite{magnanti1984network}, just to name a few.
When visually exploring networks, people often need to focus on subgraphs of interest (e.g., by selecting specific nodes and links, filtering), investigate specific connections (e.g., finding adjacent nodes, following paths), and adjust the visual properties of the network (e.g. changing graphical encodings such as color and size).
Given this multitude of tasks,   interaction plays a vital role during visual network exploration as users need to engage with both elements in the view (e.g., nodes, links) and interface controls (e.g., sliders, dropdown menus).

With the growing size and complexity of networks, recent work has begun to examine more fluid and expressive platforms and interaction techniques for visual network exploration and analysis.
Researchers have explored a number of settings including tabletops and vertical touchscreens (e.g.,\cite{frisch2009investigating,srinivasan2018orko,thompson2018tangraphe,schmidt2010set}), AR/VR (e.g.,\cite{cordeil2017immersive,hurter2018fiberclay,drogemuller2018evaluating,sun2019collaborative}), and large wall-sized displays (e.g.,\cite{kister2017grasp,lee2019dynamic,langner2018multiple}), among others, facilitating interaction through a variety of input modalities such as touch, pen, gestures, and speech.

Given the diverse and complementary strengths and weaknesses of different input modalities, an emerging theme within visualization research has been to explore multimodal interfaces that combine two or more modes of input (e.g.,~\cite{lee2018multimodal,srinivasan2018orko,kassel2018valletto,setlur2016eviza,gao2015datatone}).
In the context of network visualizations, such multimodal interfaces may enable a more fluid interaction experience~\cite{elmqvist2011fluid}, allowing people to perform common operations such as finding paths and changing visual encodings (e.g.~through speech), while simultaneously interacting with and investigating different parts of the network (e.g.~through touch).
Although multimodal interaction with network visualization seems like a promising idea, many open questions persist.
For instance, do users actually prefer multimodal input over unimodal input, and if so, why?
Does it enable them to interact more naturally, or does having multiple modes of input confuse users?
\edit{When employing multiple modalities, how do people interact with networks and perform common network visualization operations?}

We conduct a qualitative user study with 18 participants to address such questions and investigate user interactions with a multimodal network visualization system.
Ultimately, by understanding more about user interaction and preferences, we seek to help future designers build better multimodal visualization systems that are seeded by people's natural behaviors~\cite{lee2012beyond}.
Along these lines, we focus on two increasingly popular input modalities that are ubiquitous across applications and devices---namely, touch and speech.
To derive practical evidence of how people cope with system limitations and react when the system does not behave as expected, we perform this investigation using a working prototype of a speech- and touch-based network visualization tool, \textit{Orko}~\cite{srinivasan2018orko}.
We develop two unimodal (speech-only and touch-only) network visualization systems to enable comparison against the multimodal version of the Orko system.
We split participants into three groups of six participants: one group interacted with both the touch-only version and the multimodal version, the second group interacted with both the speech-only version and the multimodal version, and the third group only interacted with the multimodal version of the system.
Collectively, based on our observations and participant feedback from the study, we make the following contributions:

\begin{itemize}
    \item
    Verifying that people prefer speech- and touch-based multimodal input over unimodal input during visual network exploration, we identify specific factors explaining this preference including: the freedom of expression, the complementary nature of speech and touch, and integrated interaction experience afforded by the combination of the two modalities.
    \item Furthermore, to aid the design of future systems, we describe interaction patterns that participants employed to perform common network visualization operations (e.g., finding paths, filtering) and highlight promising areas for future work.
\end{itemize}
\section{Related Work}

\begin{figure*}[ht!]
    \centering
    \includegraphics[width=.9\textwidth]{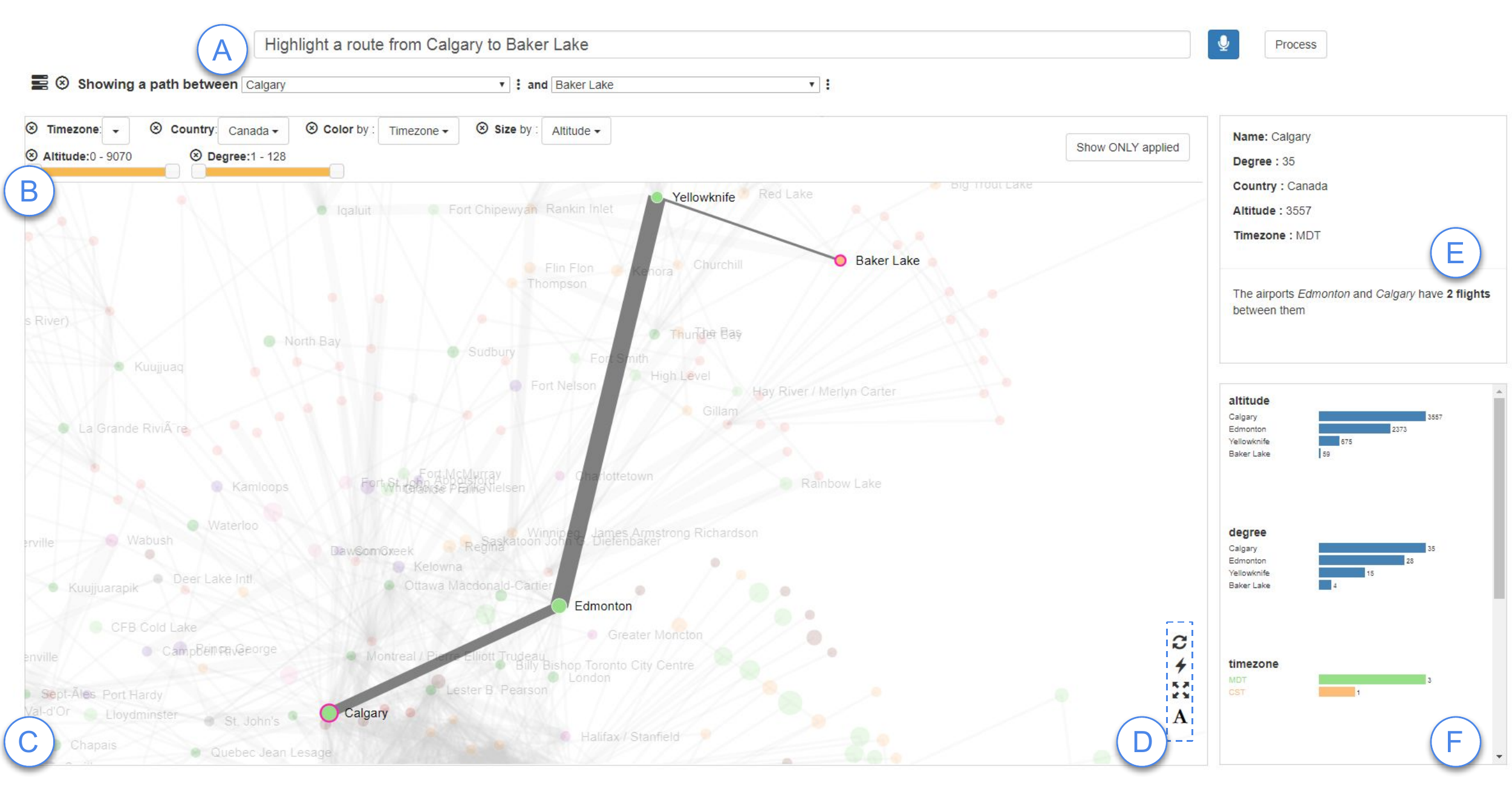}
    \caption{User interface of the multimodal study interface. A) Speech input and feedback, B) Filters and encodings, C) Visualization canvas, D) Quick-access icons, E) Details panel, and F) Summary panel. In this case, the system is highlighting a path between Calgary and Baker Lake.}
    \label{fig:system-ui}
    % \vspace{-1em}
\end{figure*}

\begin{figure}[t!]
    \centering
    % \vspace{-.5em}
    \includegraphics[width=\linewidth]{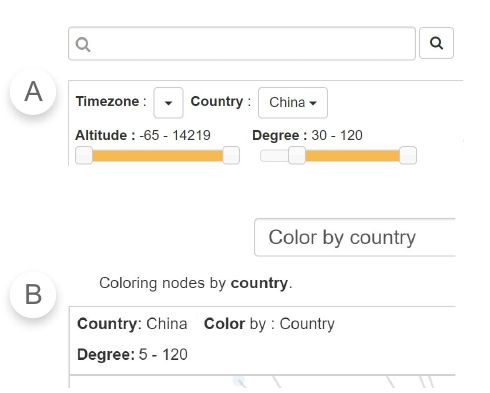}
    % \vspace{-1.75em}
    \caption{Screenshots from the unimodal study systems displaying the input, filters, and encodings rows in the (A) touch-only and (B) speech-only interface.}
    \label{fig:unimodal-ui}
    % \vspace{-1.5em}
\end{figure}

An underlying motivation for our work is based on visualization research themes highlighted in Lee et al.'s article~\cite{lee2012beyond} where the authors emphasize ``going beyond the mouse and keyboard'' as one of the key opportunities for visualization research.
Along these lines, multiple systems and studies have explored how people interact with visualizations in post-WIMP settings such as mobile/tablets~\cite{drucker2013touchviz,sadana2014designing,lee2018data,brehmer2018visualizing}, large interactive displays~\cite{lee2013sketchstory,zgraggen2014panoramicdata,langner2018multiple}, and even virtual environments~\cite{cordeil2017imaxes,wagner2018virtualdesk,hurter2018fiberclay}.
Given the widespread adoption of direct manipulation as an interaction technique in desktop-based visualization systems, a majority of these systems have explored the use of touch-based input for interaction~\cite{isenberg2009collaborative,drucker2013touchviz,baur2012touchwave,lee2013sketchstory,zgraggen2014panoramicdata,sadana2014designing,sadana2016designing}.
Furthermore, based on its increasingly important role as part of our daily interaction with technology, natural language is another form of input that has gained increased interest within the visualization research community~\cite{cox2001multi,sun2010articulate,gao2015datatone,setlur2016eviza,aurisano2016articulate2,hoque2018applying,yu2019flowsense} and as part of commercial systems~\cite{IBMWatson,mspowerbi,tableauaskdata}.
% in recent years.

While existing systems have demonstrated that both touch and natural language exhibit potential to facilitate interaction with visualizations, recent work has conjectured that the combination of the two is perhaps even more promising~\cite{lee2018multimodal}.
For instance, with Orko, Srinivasan and Stasko~\cite{srinivasan2018orko} demonstrated how speech- and touch-based multimodal interaction can be used during visual network exploration and analysis.
Kassel and Rohs~\cite{kassel2018valletto} recently presented a tablet-based visualization system, Valletto, that allows users to specify visualizations through a combination of touch and speech-based input.
% Although these systems demonstrate that multimodal input combining speech and touch is feasible for visualization, they do not provide evidence that users actually prefer such multimodal interaction over just touch- or speech-based interaction.
The development of these systems and their preliminary studies demonstrate that multimodal input combining speech and touch is feasible for interacting with visualization systems.
However, to ensure that we explore the potential of such interfaces to their fullest, we need to understand people's natural behavior, preferences, and expectations from such interfaces~\cite{lee2012beyond}.
% However, since the systems are not used to compare multimodal input to its unimodal counterparts, we lack an empirical understanding of what factors make multimodal interaction for data visualization promising or lead to its preference over unimodal input.
% As highlighted by Lee et al.~\cite{lee2012beyond}, while designing systems to demonstrate the feasibility of an interaction technique is important, understanding people's behavior and preferences is critical to create better interaction experiences with post-WIMP visualization systems.
Prior studies have explored how people use pen and touch (e.g.,~\cite{frisch2009investigating,walny2012understanding})\edit{, touch and proximity/spatial movement (e.g.,~\cite{buschel2017investigating,langner2018multiple}), and touch and tangible objects (e.g.,~\cite{besanccon2016hybrid,besancon2019HTT}), among others} in the context of visualization tools.
% For instance, Walny et al.~\cite{walny2012understanding} conducted a Wizard-of-Oz study to understand how people use pen and touch to create and modify visualizations on interactive whiteboards.
Earlier studies in the broader HCI community also have investigated multimodal interaction involving speech~\cite{pausch1991empirical,karl1993speech,oviatt1997integration} to better understand the benefits of multimodal input in terms of user performance metrics such as time and error.
% For instance, Pausch and Leatherby~\cite{pausch1991empirical} found that the combination of mouse- and speech-based input was over 21\% faster than just mouse input for graphics editing tasks.
% In other work, Oviatt et al.~\cite{oviatt1997multimodal,oviatt1997integration} compared pen and speech input to multimodal input combining them.
% They found that multimodal interaction only sped up task completion by small amounts (10\%) but significantly improved error handling and reliability (36\%).
As part of our work, we focus on exploring what aspects of speech- and touch-based multimodal input makes it promising for interaction with network visualization tools.
% We conduct a qualitative study using a prototype network visualization tool to observe how people interact using speech and touch.
% Specifically, we focus on understanding if and why people prefer multimodal input, if familiarity with an input modality changes participant behavior, and identifying key interaction patterns.

% Furthermore, to confirm that people prefer multimodal input over unimodal input and better understand reasons for this preference, we also compare their interaction with the multimodal version of the system to its unimodal counterparts.

% To address this gap, along the lines of prior studies that explore the use of pen and touch in visualization systems~\cite{frisch2009investigating,walny2012understanding,jo2017touchpivot}, and studies in the borader HCI community that explore multimodal interaction involving speech~\cite{pausch1991empirical,oviatt1997integration}, we conduct a qualitative study that investigates the role of speech- and touch-based interaction with a network visualization system.
% In addition to exploring how people interact with a multimodal network visualization system, we also compare how people interact with the system when it supports only one of the two modalities to a scenario where both speech and touch are available for interaction.
% For instance, there are many visualization systems that have explored how pen and touch can be used to facilitate multimodal interaction~\cite{}.

% \arjun{As stated earlier,} we perform our investigation in the context of a network visualization tool, Orko~\cite{srinivasan2018orko}.
Network visualizations have been extensively studied by the visualization community and many existing systems allow people to interactively explore networks by visualizing them using different layouts and representations.
A complete review of network visualization systems
% is out of scope of this paper, but
can be found in survey reports such as~\cite{herman2000graph,von2011visual,beck2017taxonomy}.
More relevant to our work, however, are taxonomies that characterize key analytic tasks and operations people perform when interacting with network visualizations~\cite{lee2006task, pretorius2014tasks, saket2014group}.
Specifically, we leverage these taxonomies to generate tasks for our user study so they are representative of what people might do when conducting visual network exploration and analysis in a realistic scenario.
Furthermore, given the crucial role of interaction while visually exploring network-based data, many researchers have examined the use of different input modalities for interacting with network visualizations across a range of devices.
For instance, Frisch et al.~\cite{frisch2009investigating} demonstrated how people use pen and touch to edit node-link diagrams on tabletops.
Schmidt et al.~\cite{schmidt2010set} and Thompson et al.~\cite{thompson2018tangraphe} also have explored multi-touch interactions for facilitating interaction with network visualizations focusing on operations like selection and basic layout editing.
More recent work has also begun to investigate gesture-based interaction with network visualizations in virtual reality~\cite{cordeil2017immersive,hurter2018fiberclay,drogemuller2018evaluating}.
Our findings contribute to this growing space of network visualization systems in post-WIMP settings by furthering our understanding of how people interact with networks using touch and speech. 
\section{Study Systems}

We used the Orko system~\cite{srinivasan2018orko} as our test bed for the study. A detailed description of Orko's features and implementation can be found in~\cite{srinivasan2018orko}\footnote{We also provide a link to videos demonstrating the Orko system as supplementary material.}.
Below we summarize the changes we made to the system for our study.
These changes were based on Orko's preliminary user study findings~\cite{srinivasan2018orko} and a series of eight pilot sessions we conducted as part of our work.

\subsection{Interfaces and Operations}

\noindent\textbf{Multimodal interface.} Figure~\ref{fig:system-ui} shows the final multimodal interface we used for our study.
Similar to the original system, the interface consists of the speech input and feedback row (Figure~\ref{fig:system-ui}A), the visualization canvas (Figure~\ref{fig:system-ui}C), quick-access icons (Figure~\ref{fig:system-ui}D), and details and summary panels (Figure~\ref{fig:system-ui}E,F).
To enhance its visibility, we repositioned the filters and encoding row (Figure~\ref{fig:system-ui}B) placing it above the visualization canvas as opposed to below the canvas in the original system.
% Users could choose to display only the applied filters and encodings similar to recent natural language-based visualization tools~\cite{gao2015datatone,setlur2016eviza,hoque2018applying} or choose to display all available filtering and visual encoding controls similar to conventional WIMP-based visualization tools.

% \begin{table}[t!]
%     \centering
%     \includegraphics[width=\linewidth]{tables/eurovis-19/operations.pdf}
%     \caption{Supported operations along with touch interactions and sample speech commands used to execute them. \arjun{change to three column T, S, MM and remove select}}
%     \label{tbl:operations}
% \end{table}

To enable comparison between unimodal and multimodal input, we developed two unimodal systems mimicking Orko's user interface components.
Table~\ref{tbl:operations} lists the operations supported across all three interfaces and how they could be performed using touch, speech, or a combination of the two.
In addition to the operations listed in Table~\ref{tbl:operations}, participants could also select nodes by tapping, drawing a lasso, or using speech (e.g.~``\textit{Select all airports in China}").

% \usepackage{booktabs}
% \usepackage{graphicx}
\begin{table*}[t!]
\centering
\caption{Operations supported in the study interfaces with their corresponding touch-only, speech-only, and multimodal interactions. Note that the speech commands shown are only examples and the systems supported a wider variety of phrasings. Speech commands preceded by \textit{`` \textgreater''} are examples of follow-up commands.}
\resizebox{\textwidth}{!}{%
\begin{tabular}{@{}llll@{}}
\toprule
\textbf{Operation} & \textbf{\touchicon{}Touch} & \textbf{\speechicon{}~Speech} & \textbf{\touchicon{}+~\speechicon{}~Touch+Speech} \\ \midrule
Find Nodes & \begin{tabular}[c]{@{}l@{}}Type label using\\ virtual keyboard\end{tabular} & \textit{\begin{tabular}[c]{@{}l@{}}“Find Calgary airport”, “Show Canberra”,\\ “Search for Auckland”\end{tabular}} & \begin{tabular}[c]{@{}l@{}}\speechicon{} \textit{“Find Canberra”} + \touchicon{}update node\\ using query manipulation widgets (Figure~\ref{fig:system-ui}A)\end{tabular} \\
 &  &  &  \\
Find Connections & Double tap on a node & \textit{\begin{tabular}[c]{@{}l@{}}“Show connections of Adelaide” \textgreater “How about Auckland”,\\ “Find airports with connections to Wales”\end{tabular}} & \begin{tabular}[c]{@{}l@{}}\touchicon{}Select nodes + \speechicon{} \textit{``Show connections"},\\ \speechicon{} \textit{``Show airports connecting to Auckland"}\\ + \touchicon{}Update node using query manipulation widgets\end{tabular} \\
 &  &  &  \\
Find Paths & \begin{tabular}[c]{@{}l@{}}Long press on source node and tap on\\ target node\end{tabular} & \textit{\begin{tabular}[c]{@{}l@{}}“Highlight a route from Normanton to Julia Creek”,\\ “Find a path between Billings and Denver”\end{tabular}} & \begin{tabular}[c]{@{}l@{}}\touchicon{}Select nodes + \speechicon{} \textit{``Highlight path"},\\ \speechicon{} \textit{``Show route between Calgary and Baker Lake"}\\ + \touchicon{}Update nodes using query manipulation widgets\end{tabular} \\
 &  &  &  \\
Filter Nodes & \begin{tabular}[c]{@{}l@{}}Adjust dropdowns and sliders\\ in filters \& encodings row (Figure~\ref{fig:system-ui}A)\end{tabular} & \textit{\begin{tabular}[c]{@{}l@{}}“Just show airports in Central Standard Time”,\\ “Filter to show Canadian airports at an altitude of over\\ 5000 feet” \textgreater “Focus on ones with degree more than 10”\end{tabular}} & \begin{tabular}[c]{@{}l@{}}\speechicon{} \textit{``Filter by degree"} + \touchicon{}Adjust degree slider,\\ \speechicon{} \textit{``Only highlight airports in Australia"}\\ + \touchicon{}Change country using dropdown\end{tabular} \\
 &  &  &  \\
\begin{tabular}[c]{@{}l@{}}Change Visual\\ Encodings\end{tabular} & \begin{tabular}[c]{@{}l@{}}Adjust dropdowns in\\ filters \& encodings row\end{tabular} & \textit{\begin{tabular}[c]{@{}l@{}}“Color airports by timezone”,\\ “Size nodes by degree” \textgreater “Now by altitude”\end{tabular}} & \begin{tabular}[c]{@{}l@{}}\speechicon{} \textit{``Color by country"} + \touchicon{}Change country\\ using color dropdown, \speechicon{} \textit{``Resize nodes"} + \touchicon{}Select\\ attribute from dropdown\end{tabular} \\
 &  &  &  \\
Navigate & \begin{tabular}[c]{@{}l@{}}Two-finger pinch for zoom,\\ one-finger drag on canvas for pan\end{tabular} & \textit{\begin{tabular}[c]{@{}l@{}}“Zoom out”, “Center graph”, “Zoom in more”, \\ “Pan left” \textgreater “Some more”, “Move right”\end{tabular}} & \textbf{---} \textcolor{gray}{(only supported through touch \textit{or} speech)} \\
 &  &  &  \\
Interface Actions & Tap quick-access icons (Figure~\ref{fig:system-ui}D) & \textit{“Refresh canvas”, “Show all node labels”} & \touchicon{}Select nodes + \speechicon{} \textit{``Show labels"} \\ \bottomrule
\end{tabular}%
}
% \vspace{.05em}
\label{tbl:operations}
% \vspace{-1.5em}
\end{table*}

\vspace{.5em}
\noindent\textbf{Touch-only interface.}
This interface functioned comparably to current network visualization tools, allowing people to select nodes and links and navigate the view with simple touch gestures, and adjust sliders and dropdown menus to filter points or change visual encodings.
Additionally, we replaced the speech input and feedback row (Figure~\ref{fig:system-ui}A) by a single input box that facilitated searching for nodes (Figure~\ref{fig:unimodal-ui}A).
Entering search terms was supported through a virtual keyboard since this was a touch-only system.

\vspace{.5em}
\noindent\textbf{Speech-only interface.}
This interface supported all operations listed in Table~\ref{tbl:operations} through speech alone.
Touch input was disabled throughout the interface.
Visually, this interface had the same components as shown in Figure~\ref{fig:system-ui} with one key difference:
since it was a voice-based system, participants could not directly manipulate the sliders and dropdown menus (Figure~\ref{fig:unimodal-ui}B).
In other words, to adjust filters or apply visual encodings, participants always had to use voice commands (e.g., ``\textit{Change the timezone to CST}," ``\textit{Size by degree}").

% For the Speech-only version of the system, the system still provided visual feedback to the users regarding available filters and the state of the system through the same kinds of interface elements. The only differentiating factor between the speech-only and the touch only-version was that in the touch-only version, users could directly manipulate these interface elements to interact with the system whereas in the speech-only version, users could only use spoken commands to achieve the same tasks. 
% \aysh{In order to start at a common ground, we made sure that both the unimodal systems were at functional parity with each other and any perceived differences in effort to achieve the actions was due to the characteristic of the modalities themselves and not due to how these systems were implemented.}

%Other than the difference in supported input modalities, the interface elements and interactions were kept consistent across all three version of the system.

% \begin{table}[h!]
%     \centering
%     \includegraphics[width=\linewidth]{tables/operations.pdf}
%     \caption{Supported operations \arjun{(TBU)}.}
%     \label{tbl:operations}
% \end{table}

\subsection{Triggering and Interpreting Speech Commands}

In the speech-only interface, participants could trigger speech recognition using the wake word ``\textit{System}'' (similar to ``\textit{Alexa},'' ``\textit{Ok Google}"). In the multimodal interface, in addition to using the wake word, participants could also tap the microphone button ({\small{\faMicrophone}}) next to the input box (Figure~\ref{fig:system-ui}A) to trigger speech recognition.
Similar to the original Orko system, all interfaces were implemented as web-based systems using Python, HTML, CSS, and JavaScript.
The standard HTML5 webkit speech recognition API~\cite{webspeechapi} was used to recognize speech input.
To improve recognition accuracy, we also trained the recognizer with system keywords (e.g., `find,' `color,' `path') and values in the loaded dataset.

We reused Orko's command interpreter~\cite{srinivasan2018orko} to respond to speech commands.
At a high-level, the system employs a two-step process to interpret commands.
First, an input command is matched against the pre-defined grammar patterns (e.g. \texttt{Color by [\underline{color}]}) defined using the Artificial Intelligence Markup Language (AIML)~\cite{bush2001artificial} to identify the operations (e.g., \textit{Find Path}, \textit{Filter}) and attributes/values (e.g., \textit{Calgary}, \textit{altitude}).
If the input command does not match a pre-defined pattern, the system then tokenizes the command string and compares it to the underlying lexicon \edit{(composed of attributes and values in the dataset, as well as keywords such as `find,' `path,' `filter,' etc.)
Based on this comparison, the system identifies} the operations and attributes/values using both syntactic (cosine similarity~\cite{singhal2001modern}) and semantic (Wu-Palmer similarity score~\cite{wu1994verbs}) similarity metrics.
While we preserved the underlying architecture, to design the speech-only interface and support equivalence between interfaces, we extended the grammar and lexicon to support navigation operations (zoom/pan) through speech (e.g., ``\textit{zoom in}," ``\textit{pan left}").
% \begin{table*}[ht!]
%     \centering
%     \caption{\arjun{Sample Tasks used in the study.}}
%     \includegraphics[width=\textwidth]{tables/tasks-subset.pdf}
%     \label{tbl:tasks}
% \end{table*}

% \input{tables/tasks-colored.tex}
\begin{table*}[t!]
\centering
\caption{\edit{Tasks used in the study.}}
\label{tbl:tasks}
\resizebox{\textwidth}{!}{%
\begin{tabular}{@{}cll@{}}
\toprule
{\color[HTML]{000000} \textbf{Task}} & \multicolumn{1}{c}{{\color[HTML]{000000} \textbf{Unimodal (Asia Pacific Flight Network)}}} & \multicolumn{1}{c}{{\color[HTML]{000000} \textbf{Multimodal (US-Canada Flight Network)}}} \\ \midrule
{\color[HTML]{000000} } & {\color[HTML]{000000} } & {\color[HTML]{000000} } \\
\multirow{-2}{*}{{\color[HTML]{000000} \textbf{T1}}} & \multirow{-2}{*}{{\color[HTML]{000000} \begin{tabular}[c]{@{}l@{}}Which of these airports have direct flights to both Auckland and Canberra:\\ {[}Melbourne, Perth, Townsville, Adelaide, Queenstown{]}.\end{tabular}}} & \multirow{-2}{*}{{\color[HTML]{000000} \begin{tabular}[c]{@{}l@{}}Consider only one hop journeys from Hartsfield Jackson to Ted Stevens airport.\\ Show that there are exactly 8 possible layover airports.\end{tabular}}} \\
{\color[HTML]{000000} } & {\color[HTML]{000000} } & {\color[HTML]{000000} } \\
{\color[HTML]{000000} \textbf{T2}} & {\color[HTML]{000000} \begin{tabular}[c]{@{}l@{}}Consider Auckland airport and the airports it has direct flights to:\\ - Among these airports, show that Auckland has most flights to Sydney Kingsford\\ Smith.\\ - Show that among the Chinese airports that Auckland has direct flights to,\\ Beijing Capital is the busiest airport.\\ - Now assume you had to fly to Western Australia from Sydney Kingsford Smith airport.\\ Name the most accessible airport in Western Australia that Sydney has a non-stop flight to.\end{tabular}} & {\color[HTML]{000000} \begin{tabular}[c]{@{}l@{}}Consider Edmonton airport and the airports it has direct flights to:\\ - Of all these airports, show that Yellowknife is the airport it has the most flights to.\\ - Consider airports in the United States that Edmonton has direct flights to. Show that\\ Palm Springs is the least busy airport.\\ - Now assume you had to fly from Edmonton to a city in the Central US region and\\ you are traveling through Palm Springs. Name the airport in the Central US region\\ that is most accessible from Palm Springs.\end{tabular}} \\
{\color[HTML]{000000} } & {\color[HTML]{000000} } & {\color[HTML]{000000} } \\
{\color[HTML]{000000} \textbf{T3}} & {\color[HTML]{000000} \begin{tabular}[c]{@{}l@{}}Let us call airports that have direct flights to 55 or more airports as “popular” airports.\\ - T/F: China has the most number of “popular” airports.\\ - Now assume that you are traveling from Sydney Kingsford Smith airport to\\ Domodedovo through one of these “popular” airports.\\ T/F: you have to travel through either Thailand or China.\end{tabular}} & {\color[HTML]{000000} \begin{tabular}[c]{@{}l@{}}Show that there are only two Canadian airports that have direct flights to 40 or more\\ other airports.\\ T/F: Among all airports that have direct flights to both these Canadian airports,\\ Denver is at the highest altitude.\end{tabular}} \\
{\color[HTML]{000000} } & {\color[HTML]{000000} } & {\color[HTML]{000000} } \\
{\color[HTML]{000000} \textbf{T4}} & {\color[HTML]{000000} \begin{tabular}[c]{@{}l@{}}List the airports you would have to fly through when travelling from Normanton\\ to Julia Creek.\end{tabular}} & {\color[HTML]{000000} \begin{tabular}[c]{@{}l@{}}Suppose you want to fly from Fairbanks to Wales. Find a set of airports through\\ which you must fly.\end{tabular}} \\
{\color[HTML]{000000} } & {\color[HTML]{000000} } & {\color[HTML]{000000} } \\
{\color[HTML]{000000} \textbf{T5}} & {\color[HTML]{000000} \begin{tabular}[c]{@{}l@{}}Assume you live in Brisbane and you want to go a high altitude location (\textgreater{}2100\\ feet) in Australia. Since there are are no direct flights to such locations, you would\\ have to travel through at least one other airport when travelling from Brisbane.\\ Name airport(s) you could fly through.\end{tabular}} & {\color[HTML]{000000} \begin{tabular}[c]{@{}l@{}}Say you are living in Billings and you want to go for a vacation to a high altitude\\ location (\textgreater{}7000 feet). However, Billings Logan does not have direct flights to\\ such locations, but it has a direct flight to an airport that does. Name that airport.\end{tabular}} \\
{\color[HTML]{000000} } & {\color[HTML]{000000} } & {\color[HTML]{000000} } \\
{\color[HTML]{000000} \textbf{T6}} & {\color[HTML]{000000} \begin{tabular}[c]{@{}l@{}}Pick any two airports that have at least one direct international flight. Consider\\ these two airports and the airports they have direct flights to. Now compare the\\ two groups of airports with respect to different characteristics such as accessibility,\\ altitude levels, variability in time zones, etc. You may also list any additional\\ observations you make based on interacting with the network.\end{tabular}} & {\color[HTML]{000000} \begin{tabular}[c]{@{}l@{}}Pick any two airports that have at least one direct international flight. Consider\\ these two airports and the airports they have direct flights to. Now compare the\\ two groups of airports with respect to different characteristics such as accessibility,\\ altitude levels, variability in time zones, etc. You may also list any additional\\ observations you make based on interacting with the network.\end{tabular}} \\ \bottomrule
\end{tabular}%
}
\end{table*}

\section{Study}

The ultimate objective of our study was to understand how people interact with network visualization tools using touch, speech, and a combination of the two.
More specifically, we had three key goals when conducting the user study in the context of a network visualization tool:

\vspace{.5em}
\textbf{RG1} Understand if and why multimodal interaction is preferred over unimodal interaction.

\vspace{.5em}
\textbf{RG2} Understand if and how prior experience of interacting using one input modality impacts subsequent multimodal interaction.

\vspace{.5em}
\textbf{RG3} Identify different input and interaction patterns people use for common operations during visual network exploration.

\subsection{Methodology}

We conducted a qualitative study where two groups of participants first interacted with either a speech-only or a touch-only interface followed by the multimodal interface.
This allowed us to collect participant preferences and feedback to compare unimodal and multimodal interaction (\textbf{RG1}).
As a baseline to see how people interact with the multimodal interface when they encounter it for the first time (without having worked with the speech- or touch-only version), we also included a third group of participants who only interacted with the multimodal version of Orko.
Comparing the interactions of the first two groups with the third group allowed us to check if prior experience using the system with just one of the modalities resulted in any notable differences in terms of interaction behavior (\textbf{RG2}).

We considered different study designs including a three condition (touch, speech, multimodal) within-subjects study and a study where participants used unimodal touch or speech input and multimodal input in counterbalanced orders.
However, a within-subjects study with three conditions would last over three hours ($\sim$60 min.~per condition) which was impractical. 
In the second alternative, having participants use the unimodal system after the multimodal system would not allow us to assess the priming effects of an individual modality.
In other words, if participants interacted with the multimodal interface first, they would already have experienced all the supported interactions, not allowing for any assessment based on prior experience using individual modalities.

\subsection{Participants and Experimental Setup}

We recruited 18 participants (P1-P18), ages 18-66, five females and 13 males.
14 participants were native English speakers and the remaining four participants self-reported as being fluent English speakers.
We sent recruitment emails to university mailing lists and recruited participants on a ``first come first serve'' basis.
Participants who only interacted with the multimodal system (P13-P18) received a \$10 Amazon Gift Card as compensation whereas participants who interacted with both the unimodal and multimodal systems (P1-P12) received a \$20 Amazon Gift Card as compensation.

In terms of their backgrounds, only eight participants said they had some prior experience of working with network visualization tools but 14/18 participants (except P2, P7, P8, P9) had some experience working with general visualization tools (e.g., Tableau, Excel).
All participants had prior experience working with touch-based devices including phones, tablets, and laptops.
All but two participants (P2, P14) said they used speech-based systems (e.g., Siri, Alexa) frequently.
None of the participants had any prior experience working with touch- or speech-based visualization systems.
All participants interacted with the system running on Google's Chrome browser on a 55'' Microsoft Perceptive Pixels device.
The screen was set to a resolution of 1920 x 1080 pixels.

\subsection{Dataset and Tasks}

As the primary focus of our study was understanding user interactions, we had to ensure that the network selected for the study encouraged interaction with the visualization and allowed us to cover a wide variety of tasks including browsing, attribute-based filtering and reconfiguration, and group-level exploration~\cite{lee2006task,saket2014group,yoghourdjian2018exploring}.
Given this high-level goal, we wanted to select a dataset where: (1) nodes had both numerical and categorical attributes so participants could filter and change visual encodings and (2) the connections had an intrinsic meaning so the tasks could emulate real-world scenarios.
Additionally, to avoid differences due to domain knowledge, we wanted a dataset from a domain that was familiar to all participants (i.e., participants knew what the different attributes meant).
With these criteria in mind, we selected two undirected flight networks as our datsets for the study.

The first dataset contained 551 airports (nodes) in the Asia Pacific region and 2263 bidirectional flights between airports (links) whereas
the second dataset contained 556 airports in United States and Canada and 2219 flights between those airports.
Each airport in the dataset had four attributes including its \textit{altitude}, \textit{country}, \textit{timezone}, and a derived attribute indicating number of airports it was connected to (\textit{degree}).
Participants explored the Asia Pacific network in the unimodal condition, and the US-Canada network in the multimodal condition.

Participants performed six tasks (T1-T6) with each dataset.
\edit{Table~\ref{tbl:tasks} lists the tasks used during the study.}
These study tasks were generated based on existing network visualization task taxonomies~\cite{lee2006task,saket2014group,pretorius2014tasks} and included topology-level tasks, attribute-level tasks, browsing-tasks, and a group-level comparison task.
\edit{For instance, in terms of Lee et al.'s taxonomy~\cite{lee2006task}, T1 and T4 correspond to topology-based tasks (finding paths and connections), while T2, T3, and T5 involve a combination attribute-based tasks (filtering), browsing (following paths), and topology-based tasks.}
For P1-P12, tasks between the unimodal and multimodal conditions were designed such that they had a comparable level of difficulty.
The order of tasks was randomized between conditions for each participant to prevent them from memorizing the operations they performed for a task.

% Table~\ref{tbl:tasks} lists the tasks used during the study.
To prevent participants from reading out the tasks as commands into the system as-is, we framed the tasks as a combination of scenario-based questions and jeopardy-style facts~\cite{gao2015datatone} that participants had to prove true/false.
In other words, to ``solve'' a task, participants had to interact with the system and get to a point where the visualization either proved or disproved the given statement or highlighted the required sub-graph.
% For instance, consider T4 in the multimodal condition (Table~\ref{tbl:tasks}).
% was ``\textit{Suppose you want to fly from Fairbanks to Wales. Find a set of airports through which you must fly.}''
For instance\edit{, consider the task}\remove{ one of the tasks for the US-Canada network was} ``\textit{Suppose you want to fly from Fairbanks to Wales. Find a set of airports through which you must fly.}'' \edit{in Table~\ref{tbl:tasks}.}
To solve this task, participants could either find the path between the Fairbanks and Wales airports or they could manually, incrementally explore connections out of one of these airports until they reached the other.
In either case, since there were multiple correct answers, participants had to visually highlight or show the list of airports (path) that one would need to travel through.

\subsection{Procedure}

Sessions lasted between 50-60 minutes for participants who only interacted with the multimodal system and 125-135 minutes for participants who interacted with both the unimodal and multimodal systems.
The study procedure was as follows:

\vspace{.25em}
\noindent{\textbf{Consent and Background} (3-5 min)}: Participants signed a consent form and answered a questionnaire describing their background with visualization tools and touch- and speech-based applications.

\vspace{.5em}
\noindent{\textbf{System Introduction} (3-5 min)}: The experimenter introduced the system, describing the user interface and supported operations.
\remove{Specifically for the}\edit{For} speech interaction, participants were only informed about the operations the system supported and were not given a detailed vocabulary or list of possible commands for each operation.
Instead, participants were encouraged to interact with the systems as naturally as possible, using any commands they felt were appropriate in the context of the given datasets.

\vspace{.5em}
\noindent{\textbf{Practice} (3-5 min)}: Participants
tested the touch and speech input until they felt comfortable using them. In this phase, participants interacted with a network of 552 European soccer players (nodes) that were linked if they played for the same club or national team (6772 links). Each node had five attributes indicating the player salary, goals scored, field position, club, and country they represented.

\vspace{.5em}
\noindent{\textbf{Dataset and Task Introduction} (3-5 min)}: Participants were given a description about the flight network dataset along with the six tasks printed on a sheet of paper.

\vspace{.5em}
\noindent{\textbf{Task Solving} (30 min)}: Participants interacted with the system to solve the tasks. This phase was capped at 30 minutes. Participants were encouraged \edit{(but not mandated)} to think aloud and interact with the experimenter\remove{while performing the given tasks}\edit{, particularly when they felt the system functioned unexpectedly. To avoid prompting interactions or disrupting the participants' workflow, the experimenter did not intervene during the session and only responded when participants initiated the discussion.}

\vspace{.5em}
\noindent{\textbf{Debrief} (10-15 min)}: Participants filled out a post-session questionnaire and engaged in an interview describing their experience with the system.

Participants who performed tasks with two systems (P1-P12) were given a 15 minute break between the two sessions.
After this break, except for the consent and background step, we followed the same procedure as with the first system. 
These participants were also asked to state and describe their preference between the unimodal and multimodal versions of the system during the debrief.
We video recorded all participant interactions with the system and audio recorded all interviews.

\subsection{Data Analysis}

Two experimenters individually reviewed both the audio and video data collected during the study to identify themes in interaction patterns and participant feedback.
The resulting themes were then collectively discussed and iteratively refined into groups of observations using an affinity diagramming approach.
This helped us characterize subjective feedback and participant behavior to qualitatively answer the initial questions driving the study (\textbf{RG1}, \textbf{RG2}).
Furthermore, we also performed closed coding of the session videos to categorize the different types of interactions performed during the study (\textbf{RG2}, \textbf{RG3}).
For this analysis, we used the operations in Table~\ref{tbl:operations} as our set of pre-established codes.
For each attempt at performing an operation, we noted if a participant used speech, touch, or a combination of the two.
For instance, if a participant filtered nodes using a single spoken query (e.g., ``\textit{Show airports located at over 2100 feet}''), we would count this as one speech-only interaction.
Alternatively, to filter, one could also directly adjust the slider (touch-only) or use a combination of the two modalities (``\textit{Filter by altitude}'' + drag slider).
The intended operations were generally apparent due to the `think aloud' protocol, the design of the tasks, and by the participant's reaction to system's interpretation of their interaction.
The closed coding was also performed by two experimenters individually and conflicting observations or mismatches in counts were collectively resolved.
\edit{We also used the session videos to determine the task completion times.}

\begin{table}[t!]
    \centering
    \caption{Distribution of 945 interactions used to perform six common network visualization operations during the study.
    \textbf{U}: Unimodal interface, \textbf{M}: Multimodal interface, \textbf{S}: Speech, \textbf{T}: Touch, \textbf{ST}: Multimodal interactions.
    A `-' indicates that a modality was not supported in a condition or that participants were not assigned to a condition.}
    \includegraphics[width=\linewidth]{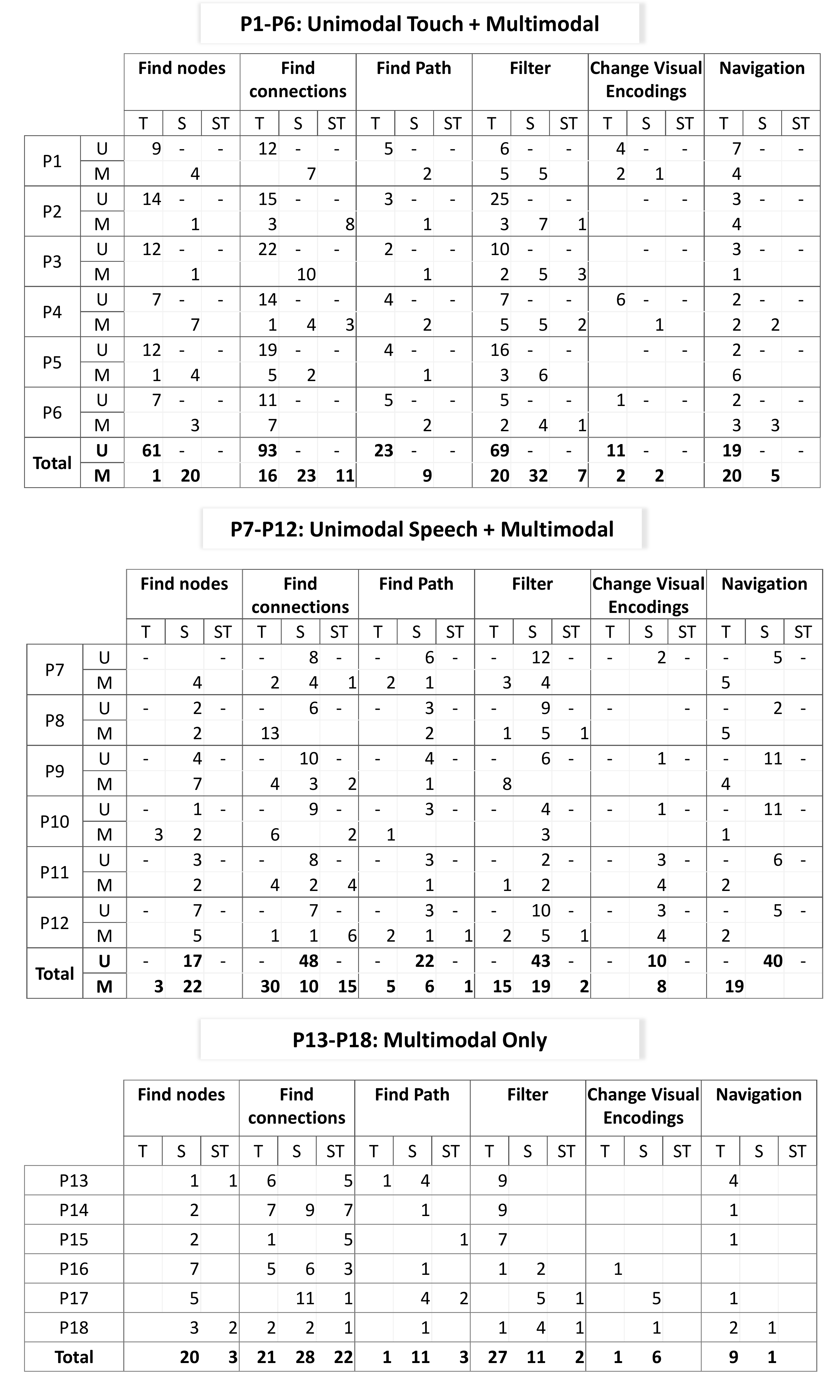}
    \label{tbl:interactions}
\end{table}
\section{Results}

% Pilots:
% \begin{itemize}
%     \item Multiselect w/ touch (check Orko paper for consistency)
%     \item Select vs Highlight (and state management in general is an open problem)
%     \item Position of filters
% \end{itemize}

% Main study:
% \begin{itemize}
%     \item Touch for reposition is important
%     \item Lack of vis knowledge limits what people try to do regardless of modalities (e.g., P3 never used color or size but P4 used colors very effectively)
%     \item Need for better ambiguity resolution techniques: one opportunity is to employ confirmation strategies (Currently system does what it thinks is right) [refer to feedback/error points and failed queries]
%     \item Touch as primary vs Speech as primary
%     \item Analysis of speech error types
%     \item Highlight special cases of speech queries (P8: remove filter by values, connections+filter, ``busiest, most number of travellers'')
% \end{itemize}

% We first reviewed the session videos and participant responses to identify groups of interaction patterns and subjective comments collected as part of the think aloud protocol and the post-session interview.
% We then used an affinity diagramming approach and iteratively refined the groups to categorize them under higher level categories that allowed us to answer the three initial questions driving the study.

Addressing our study goals, in this section, we describe our key findings corresponding to the preference for multimodal interaction (Section~\ref{sec:multimodal-preference}), the effect of priming users with one modality (Section~\ref{sec:priming}), and the different input and interaction patterns employed by the participants (Section~\ref{sec:interaction-patterns}).
% As for the strengths and weaknesses of individual modalities, we observed that they directly contributed to preference of multimodal input and the choice of interaction for different tasks, and are discussed in the respective sections.

\subsection{Task and Interaction Overview}

% \arjun{Add task completion time and success rate summary para.}
11/12 participants (P1-P12) who interacted with the unimodal interface completed all six tasks, whereas one participant (P8) completed five.
In terms of the correctness of task responses, four participants made errors: P3 and P7 answered one of the six tasks incorrectly and P8 and P9 responded incorrectly to two tasks.
In the 18 sessions with the multimodal interface, all participants except P18 (who completed five tasks) completed all six tasks with only three participants (P3, P10, P15) making one error (each) while responding to the six study tasks.
In terms of time, the task phase lasted, on average, 24 minutes with the touch-only interface, 23 minutes with the speech-only , and \edit{21 minutes with the multimodal interface}.
% Unimodal touch: 1457.33
% Unimodal speech: 1364.33
% Multimodal: 1410.67

We recorded a total of 1052 interactions corresponding to the seven operations in Table~\ref{tbl:operations} across the 18 participants and the two study interfaces.
Table~\ref{tbl:interactions} shows the distribution of 945/1052 interactions for six operations (\textit{Find Nodes}, \textit{Find Connections}, \textit{Find Path}, \textit{Filter}, \textit{Change Visual Encodings}, \textit{Navigate}) that are common across network visualization systems.
We exclude \textit{interface actions} from Table~\ref{tbl:interactions} since these are generic tool-level operations (e.g. refreshing the canvas) and are not specific to network visualizations.

% \arjun{Need some sort of intro sentence + state that we report findings here but provide high-level takeaways in the last section.}

\subsection{Preference for multimodal interaction}
\label{sec:multimodal-preference}

When asked which of the two systems they preferred, all 12 participants (P1-P12) who worked with both the unimodal and multimodal interfaces said that they preferred the multimodal system over the unimodal system.
% However,
This was not surprising given similar findings in earlier studies~\cite{pausch1991empirical,oviatt1997integration} and the simple fact that the multimodal system provided all capabilities that the unimodal system did.
Hence, we were more interested in understanding what aspects of the combination of speech- and touch-based interaction with the system led participants to prefer it (\textbf{RG1}).

% To understand potential reasons for this preference, we first considered the participants interactions in both conditions.
% Table~\ref{tbl:interactions} summarizes the interactions for the 18 participants for each task and condition.
% The cell values indicate the number of times input modalities were used individually or in combination while performing a task.
% We recorded an interaction each time a participant issued a speech command and when they used touch or a combination of speech and touch to perform one or more operations in Table~\ref{tbl:operations}.
% For instance, in the touch-only condition, during task T1, P1 performed two interactions for \textit{find}, one for \textit{clear selection}, and two interactions for \textit{finding connections}.
% However, in the multimodal setting, P1 issued two speech commands (one failed and one successful command for \textit{finding connections} of nodes), tapped once on the canvas background to \textit{clear selections}, and used touch once to \textit{pan} the view.

% After reviewing the interaction counts,
One hypothesis for why participants preferred the multimodal system, developed after reviewing their interaction counts in Table~\ref{tbl:interactions}, was that in some cases, multimodal input allowed them to perform tasks with fewer interactions.
However, tasks could be performed using multiple strategies through varied operations, each resulting in a different number of steps.
Thus, basing the preference on interaction counts alone would be unjustified because we did not control for which strategy or operations participants used during a task.
% However, since the counts included failed attempts at operations (e.g., failed speech queries) and actions that were not necessarily required to solve a task (e.g., finding connections of a node that was not listed in the task), basing the preference on interaction counts alone would be unjustified because we did not control for such factors.
Instead, we coupled the participants' verbal comments and our observations of their interactions to identify three factors listed below that we believe led to their preference for multimodal interaction.

% \arjun{Maybe good to give numerical overviews of uni vs multi here in terms of speed (time) and operation counts.}

\subsubsection{Freedom of expression}

Out of the 945 interactions, 489 were performed in the context of the mulitmodal interface (P1-P6 M, P7-P12 M, and P13-P18 in Table~\ref{tbl:interactions}).
Among these, 233 (48\%) used unimodal speech input, 190 (39\%) involved unimodal touch input, and 66 (13\%) used both modalities sequentially.
% These results are comparable to Oviatt et al.'s seminal studies investigating the use of speech and pen-based multimodal input which highlights that users' expressed commands multimodally only 20\% of the times~\cite{oviatt1997integration}.
Although only 13\% of interactions involved sequential use of modalities, all participants used both modalities (individually or together) during at least three out of the six tasks in a session.
% (observable through the rows on the right side of Table~\ref{tbl:interactions}). 

% Input patterns also varied across participants for the same study task.
% For instance, P13 issued five speech commands and two multimodal commands when performing a task whereas P14 used one unimodal speech command, one unimodal touch action, and one multimodal command when performing the same task.
Interaction patterns also varied across participants for the same operation.
For instance, observing the interactions for P13-P18 in Table~\ref{tbl:interactions}, we can notice that P17 and P18 primarily used speech (individually or sequentially with touch) for filtering.
On the other hand, P13-P16 primarily used touch to filter nodes.
Interaction patterns varied even for individual participants across tasks.
For instance, while performing the first task, P1 issued a unimodal speech command to find connections.
However, during the second task, to find connections, he used speech and touch sequentially.

Participants also verbally commented on their preference for multimodal input over unimodal input in their post-session interviews.
Participants said that having multiple modes of input gave them more freedom to try different ways to perform a task.
For instance, highlighting the use of speech and touch for different operations, P8 said ``\textit{The combination is certainly better. Voice is great when I was asking questions or finding something I couldn't see. Touch let me directly interact.}''
Similarly, P2 said ``\textit{It (multimodal input) felt more natural. I really liked that I could choose what I wanted to do with my hands and what I wanted to say.}''
Stating multimodal interaction was more natural, P10 also said ``\textit{Working with this second system felt more natural. If I wanted to filter by something I could just say that but when I'd see something interesting I could touch it without having to say something and wait for the system to process it.}''
% Commenting on the P6 said ``\textit{I felt more free to try out different features.}'' and also commented on multimodal interaction being more natural, stating ``\textit{I think it goes to this idea that humans like the idea of having more options and the flexibility in interaction.}''

The varied interaction patterns within and across participants coupled with the subjective comments highlight how the multimodal interface provided more freedom of expression, allowing participants to interact based on the task context or personal preferences.

% These varied interaction patterns and preferences between different participants and across tasks for the same participant suggest that different modalities may be more natural or convenient given the context of a task or a participant's personal preference.
% Coupled with the subjective comments, these observations highlight how the multimodal interface accommodated these preferences.

% However, to facilitate truly flexible multimodal input, further work needs to be done to understand how visualization system users perform various tasks using modalities like speech and touch.
% Based on a combination of participant preferences, their subjective feedback, and the observed interactions, we argue that there is value in multimodal input for visualization tools in general and hence, we need to design more systems that help us investigate the role of multimodal interaction in information visualization.

\subsubsection{Complementary nature of modalities}

A popular hypothesis about multimodal interaction is that it allows users to offset the weaknesses of one modality with the strengths of another~\cite{cohen1989synergistic,lee2018multimodal}.
Along these lines, when describing their experience with the multimodal system, 12 participants (P2, P4-6, P9-11, P13, P15-18) explicitly commented on the complementary nature of touch and speech and how it was a key advantage of multimodal input.

Participants found the ability to correct speech with touch very useful, with some participants even stating that the combination is vital to make effective use of speech.
For example, P17 said ``\textit{I liked that I could correct with touch. Because it's not always going to be perfect right. Like the smart assistant on the phone sometimes gets the wrong thing but doesn't let me correct and just goes okay.}''
Talking about cases when the system populated the right filtering attribute but did not detect the right value, P2 said ``\textit{the system would bring the correct dropdown even if it didn't get the value right and then I could simply correct that.}''
% Valuing the use of touch to complement speech even further, P13 said ``\textit{Without the redundancy of touch, voice fails.}''
In addition to correcting speech recognition and ambiguity errors with touch, participants also appreciated that they could leverage touch to modify existing queries.
For instance, referring to the query manipulation dropdowns in the speech input feedback row (Figure~\ref{fig:system-ui}A) 
% that allowed changing values in a query without having to verbally repeat the query, 
, P18 said ``\textit{I liked that it allowed me to modify my command without having to say it again.}'' 

On the other hand, participants also found the ability to use natural language when they either forgot a gesture or were unable to perform an operation using touch.
P13, for instance, said ``\textit{I used voice when I didn't know how to do it with touch.}'' highlighting that speech can aid in overcoming memorability issues associated with touch gestures.
Similarly, five participants (P4, P6, P13, P15, P18) used speech commands to navigate the view (i.e., zoom and pan) when they were in a dense region of the network and were unable to pinch or drag without touching the nodes on the canvas.

% \arjun{Some high-level summary/takeaway here.}
In addition to affirming the benefits of complementary interactions, such comments also highlight the value in further exploring elements like ambiguity widgets~\cite{gao2015datatone} to help users resolve challenges with speech and query manipulation widgets~\cite{srinivasan2018orko} that help users modify existing utterances.
% about the complementary use of modalities support the popular hypothesis that multimodal interaction allows users to offset the weaknesses of one modality with the strengths of another~\cite{cohen1989synergistic,lee2018multimodal}.
% These observations also highlight value in further exploring elements like ambiguity widgets~\cite{gao2015datatone} to help users detect and resolve ambiguity with speech and query manipulation widgets~\cite{srinivasan2018orko} that help users modify existing utterances.

\subsubsection{Integrated interaction experience}

Another theme among the participants' comments regarding the advantages of multimodal interaction alluded to the notion of \textit{fluidity} as characterized by Elmqvist et al.~\cite{elmqvist2011fluid}.
% For instance, P18 described his experience as ``\textit{super natural}" stating ``\textit{I found it very natural to use multiple modes and quickly learnt to take better advantage of that.}"
For instance, referring to the ability of being able to apply filters while interacting with nodes on the view, P16 said ``\textit{Generally I prefer touch but here speech was good because then I don't have to look through filters and I can just say it while moving points.}"
Although he was initially skeptical about multimodal input, during his interview, P10 said ``\textit{It was somehow less complex even though more interactions were added.}"~suggesting that the combination of modalities helped reduce the overall cognitive load.
The comparatively fluid nature of multimodal input also led to participants perceiving themselves as being faster with the task even though the overall task completion times were comparable across the study interfaces.
For instance, P1 said ``\textit{Having the combination was a lot easier to work with. Instead of having to find nodes and then highlight connections, I could do it in one command and then continue to interact with the graph.}"
% \arjun{P18, P16 (filters comment), P10 (somehow less complex...), P8 (listen to audio at 4min), P1 (step reduction)}
% \aysh{ For instance, P8, who interacted with both the unimodal and multimodal systems said ``\textit{ You can step back.. you are spending more time looking at the data itself} " about using the multimodal system. This participant also compared using both modalities as being analogous to how he might naturally reaching for something closer to him at a dinner table or simply ask someone else to pass it on to him if it is not within his reach}

% Specifically, six participants (P2, P7, P8, P10-12) who interacted with both the unimodal and multimodal system commented on the multimodal system providing higher degrees of freedom during visual exploration.
% Commenting on the freedom of expression facilitated by multiple modalites, P7 said ``\textit{The system gave me a lot more freedom to choose how I wanted to do something but didn't overwhelm me at the same time.}''

Motivated by such comments, we further reviewed the session videos to better understand what specifically about multimodal input evoked the feeling of fluidity.
Based on our review, we attribute the fluidity of interaction in the multimodal interface to speech and touch facilitating \textit{integrated interactions}~\cite{walny2012understanding}
% Based on our observations, we hypothesize that having the option to use speech and touch facilitates \textit{integrated interactions}~\cite{walny2012understanding} which, in turn, promote fluid interaction.
that are defined as ``\textit{interactions where a person's hands, tools, actions, interactions, visual response, and feedback are in situ where the data is visualized. That is, to effect an interaction, a person's attention is not drawn away from the visual representations of data in which they are interested.}''
Examples of integrated interactions during the study included applying a filter using speech while dragging nodes---not having to take eyes off the nodes in focus, or the ability to find nodes using speech without having to divert attention in order to type on a virtual keyboard that occluded the underlying visualization, among others.
While these are seemingly straightforward interactions in isolation, they illustrate that the modalities together allowed participants to stay in the flow of their analysis rather than divert their attention to other user interface elements.

\subsection{Effects of priming users with speech or touch}
\label{sec:priming}

% In addition to understanding preferences between unimodal and multimodal input, we also wanted to investigate if users interacted differently when they used the multimodal system after having prior experience using one of the two modalities (\textbf{Q2}).
One of our study goals also was to observe if participants interacted differently with the multimodal system when they had prior experience with one of the two modalities (\textbf{RG2}).
More specifically, we were curious if participants would continue to use the same modality and not use multimodal input? Would participants rely more heavily on the modality they first experienced? Would interaction patterns for these participants notably differ from those who only interact with the multimodal system?
% Are there specific operations that people switch 
We were interested in these questions as the findings could challenge the need for multimodal input altogether.
For instance, one possible outcome was that participants who interacted with the unimodal touch system (P1-P6) would continue to use only touch in the multimodal setting and similarly P7-P12 would use only speech input in the multimodal setting.
Such a finding would suggest that people resort to what they know, refraining from learning new ways to interact with a system.
Alternatively, it could also imply that adding input modalities may have limited (or no) benefits when users know how to work with a system using a specific modality.

% We observed no notable differences between participants who used the multimodal system directly (P13-P18) and participants who had prior experience working with the unimodal system (P1-P12), however.
% We observed that participants who had prior experience working with the unimodal system (P1-P12) as well as participants who worked with the multimodal system directly (P13-P18) both used the two modalities comparably in the multimodal condition (Table~\ref{tbl:interactions}).
We observed that participants who had prior experience working with the unimodal system (P1-P12) interacted with the multimodal system comparably to participants (P13-P18) who worked only with the multimodal system.
When we explicitly asked participants if their experience with the first system affected their behavior, participants said that they used both modalities subconsciously and did not think about it until we asked them to reflect on it.
For instance, P12 said ``\textit{Now that I think of it, not consciously but I did use speech to mostly narrow down to a subset and then touch to do more detailed tasks.}"
% \arjun{P12 quote..also use as segue into operations}.
In fact, perhaps the single most important aspect that decided which modality would be used was the operation being performed.
For instance, consider participants P1-P6 and the \textit{find path} operation in Table~\ref{tbl:interactions}.
In this case, when interacting with the multimodal interface, all participants switched to using only speech commands even though they had all previously performed the operation using touch.
Furthermore, this interaction pattern of primarily using speech for finding paths is comparable to the participants in the other conditions (P7-P18), suggesting a general mapping between the modality and task.
Combined with the comments from the previous section, these observations suggest that people naturally adapted to using a new modality that was more suited for an operation even if they were experienced at performing the same operation with a different mode of input.
% Combined with previously highlighted subjective comments and preferences, these observations motivate the further investigation of multimodal visualization interfaces that provide more freedom of expression, allowing people to choose the most suited (or preferred) modality for an operation.
% \arjun{Maybe create transition table for more specific results.}

\begin{figure}[t!]
    \centering
    \includegraphics[width=\linewidth]{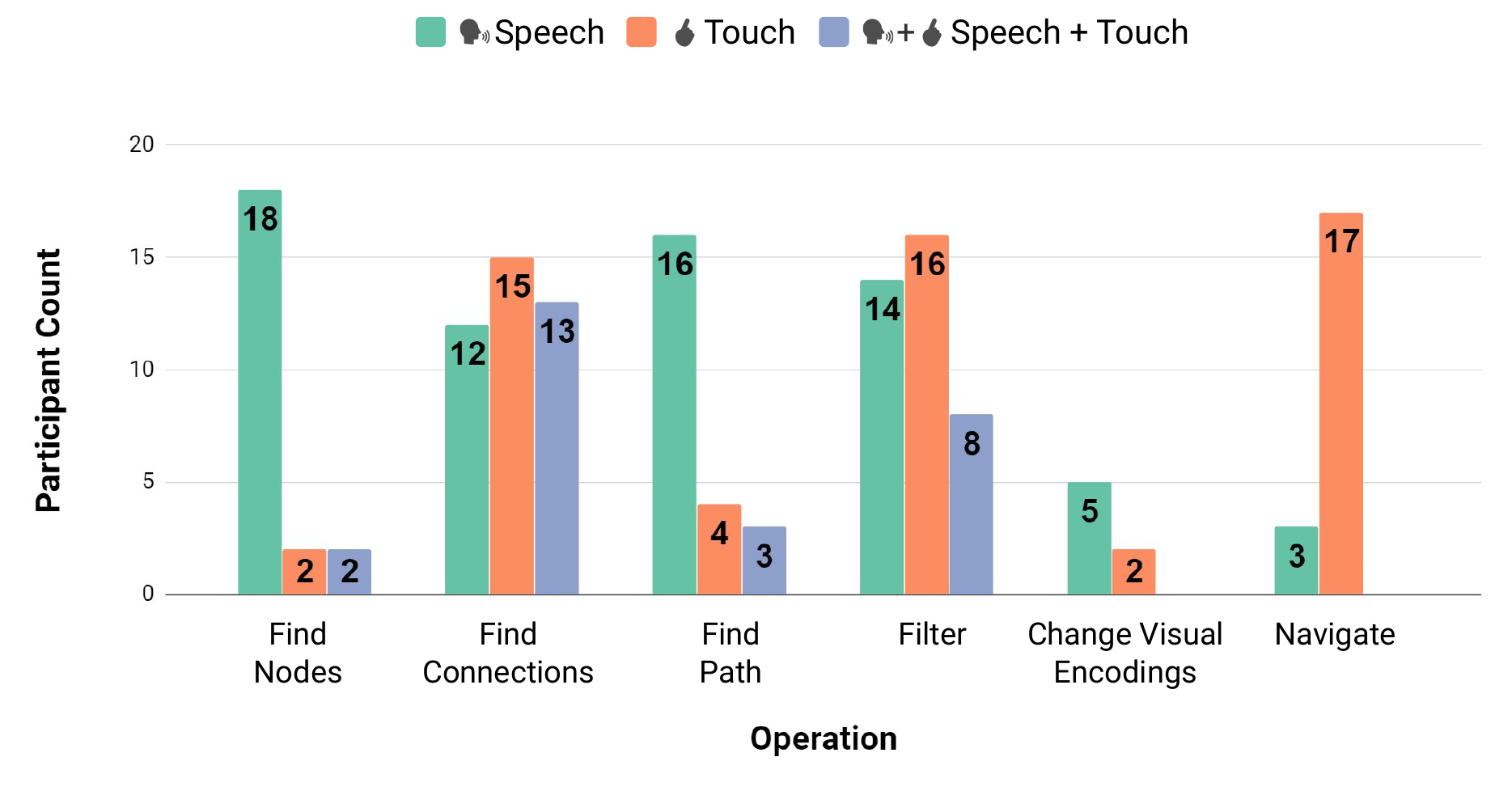}
    \caption{\edit{Number of participants using different modes of input in the multimodal interface for each type of operation.
    Navigation was primarily performed using touch, whereas finding nodes and paths was largely performed through speech.
    Other operations had more variety in input patterns.}
    }
    \label{fig:operation-input-counts}
\end{figure}

\subsection{Operations and interaction patterns}
\label{sec:interaction-patterns}

% \remove{We now highlight participants performed when executing common network visualization operations (\textbf{RG3}).}
\edit{Figure~\ref{fig:operation-input-counts} summarizes the number of participants who used speech-only (~\speechicon{}), touch-only (\touchicon{}), or combined speech and touch input (~\speechicon{} +\touchicon{}) for performing common network visualization operations (\textbf{RG3}).
Note that a single participant may have performed more than one type of interaction for an operation (e.g., as shown in Table~\ref{tbl:interactions}, for finding paths, P13 used both speech-only and touch-only interactions).}
Our goal here is not to suggest one ``best" input modality or interaction for an operation but rather to highlight the variety in patterns so future system designers can make more informed interaction design decisions.
For this analysis, we only considered the 489 interactions with the multimodal interface when listing the actions since the unimodal interfaces did not give participants the option to choose their preferred style of input.

At a first glance, both Table~\ref{tbl:interactions} and Figure~\ref{fig:operation-input-counts} suggest that participants largely performed operations using a single modality, infrequently \textit{combining} modalities.
However, it is important to also note that participants switched between modalities for different operations, using both modalities individually or together at some point during all sessions.
Affirming to the myths of multimodal interaction~\cite{oviatt1999ten}, this switching between modalities highlights that the value of the multimodal interaction does not only come from modalities being combined but also stems from the availability of different modalities to perform varied operations amidst a task.
% S: P4, P6, P18\\
% T: all 18

% \subsection{Comparing Touch and Speech \arjun{Not sure about keeping this subsection}}

% \subsubsection{WIMP Overloading}

% Refer to Maes vs Shneiderman debate \& BM\&K paper.
% Speech helps with this... DM/Touch has this issue.

% \subsubsection{Speed}

% Typically higher with speech.. even with repetition of commands... Touch can be tedious

% \subsubsection{Cost of Error}

% Higher with speech and the solution was often ``Reset''. Lack of appropriate feedback has a role in this.

% \subsubsection{Trust}

% People trusted touch more... some were even hesitant to use voice. Was unexpected to the extent we saw.. but also highlights importance of multimodal so people have a choice

% \subsubsection{Preference + Learnability}

% Most felt speech was easier to learn and was more natural. However, this was likely because of the simple nature of operations and commands used. For more complex systems with multiple chart types, we suspect this may not be as strong a case.
\section{Discussion\remove{ and Future Work}}
\label{sec:discussion}

Our observations and participants' subjective feedback during the study also guided us to some higher-level findings and takeaways that we discuss below.

\subsection{Dissecting the integration of speech and touch}

% One observation in our study was that there was no simultaneous use of modalities (i.e., there was no temporal overlap between speech and touch.)
% This is consistent with Srinivasan and Stasko's preliminary study~\cite{srinivasan2018orko} and is also in line with prior study findings that reveal that multimodal signals involving speech and pointing or gesture rarely co-occur temporally during human-computer communication~\cite{oviatt1997integration,oviatt1999ten}.

Among the interactions that used both modalities, touch preceded speech in only 1/66 cases.
This is in stark contrast to Oviatt et al.'s study investigating pen and speech-based interaction where 99\% of the sequential multimodal constructions involved the use of pen before speech~\cite{oviatt1997integration}.
However, this pattern of speech preceding touch in visualization systems was also observed in a preliminary study with Orko~\cite{srinivasan2018orko}.
Setlur et al.~\cite{setlur2016eviza} also demonstrate such interactions in Eviza with queries like ``\textit{earthquakes with magnitudes between 4 and 6 here}'' (where \textit{here} is later specified through a lasso drawn using the mouse).
We attribute this contrast between sequential use of modalities in recent studies of visualization systems to Oviatt et al.'s~\cite{oviatt1997integration} seminal study with the QuickSet system to three factors.
First, it may be a practical constraint arising from the modalities used and the study task.
Specifically, Oviatt et al.'s study focused on the task of drawing on a map and used a digital pen as one of the input modalities.
In contrast, recent studies of visualization systems (including ours) use either touch~\cite{srinivasan2018orko} or mouse~\cite{setlur2016eviza}, focusing on visual information seeking tasks.
% First, it may be a practical constraint arising from the use of touch (in~\cite{srinivasan2018orko} and our study) or mouse (in~\cite{setlur2016eviza}) instead of a pen (in~\cite{oviatt1997integration}) and the target task (drawing on a map in Oviatt et al.'s study and visual information seeking in recent studies).
Our second hypothesis is that speech interfaces have become much more popular now and people more commonly treat speech as a primary input modality.
The third reason for this behavior could be that in the context of a network visualization, the sequential integration of speech and touch best maps to Shneiderman's ``\textit{Overview first, zoom and filter, then details-on-demand}'' mantra~\cite{shneiderman1996eyes}.
In other words, people get an overview by looking at the view, use speech to filter since it affords simultaneous specification of multiple filtering criteria, and then use touch to get details-on-demand due to the precision it affords.
% \textbf{<MAYBE an interesting place to highlight contrast w/ Oviatt work and say S+T is more common in vis due to Overview first mantra>}

% \subsection{Touch-based network visualization systems are an open area for research}

% Similar to Srinivasan and Stasko's initial prototype, our system only supported a small set of touch gestures including single tap, double tap, long press, and pinch and drag.
% We assumed that participants would not have any difficulty in remembering and using these gestures since these largely map to their mouse-based counterparts (e.g., single click for select, double click to highlight connections.)
% To our surprise, however, nine participants (P1-3, P6, P9, P12-15) got confused on multiple occasions when using touch.
% The most common source of confusion was using long press for finding the path versus finding connections.
% We suspect this is because of the participants' familiarity with context menus on desktop systems that typically appear on right-click (equivalent of long press) and provide operation options.
% In other cases, the participants simply could not remember which gesture mapped to which operation (e.g., P14 kept trying to select a node through double tap.)
% While participants eventually got used to the interactions, such observations highlight that empirically designing effective and standardized touch-based interactions for network visualizations is an open research problem.
% % Even simple interactions like double-tap and long-press were confusing. Needs to be more work in layouts.

\subsection{Enhancing Discoverability}

One hypothesis emerging from an input modality (or even multimodal input) being more natural for interaction is that it encourages users to explore a wider range of operations.
We did not observe this behavior, however.
In both the unimodal and the multimodal conditions, participants largely resorted to single step and fundamental network visualization operations (find, find connections, and path).
% While filtering was also used, it was typically only when it was absolutely needed for a task.
Operations such as changing the color and size of nodes were also performed less frequently (7/18 sessions) even though they proved more effective than filtering for participants who leveraged them.
However, the participants who did explore a wider range of operations were mostly those with prior experience using network visualization tools.
While this can be attributed to the system implementation and tasks to an extent, it also highlights an important consideration regarding the system discoverability.

Discoverability applies to both aspects of discovering \textit{what} operations are supported as well as \textit{how} to perform them~\cite{norman2013design}.
% We argue that input modalities like speech and touch (and even their combination) may help with \textit{how} to perform the operations they intend to perform.
% However, they do not inherently help users think about the different methods in which similar operations can be performed (i.e., the \textit{what}).
Particularly, since the system supported speech input, there was not always a one-to-one mapping between the GUI and possible operations (e.g. dropdowns for changing visual encodings showed up only when invoked via speech), potentially resulting in participants forgetting the operation.
While the initial training and practice phase helped participants get acquainted with how to use the system, recollecting which operations could be performed during tasks was a common challenge.
In fact, realizing he could have performed some initial tasks faster had he used the \textit{find path} operation, during his interview, P13 said ``\textit{In fact, it would be helpful if I could tap on the nodes and the system could remind me of what I could do}." \edit{hinting at the use of feedforward techniques~\cite{norman2013design,vermeulen2013crossing} to aid discoverability of speech input.
To this end, one idea for future systems to explore may be to suggest contextually-relevant operations and corresponding speech commands based on the active state of the view and previously performed interactions (e.g.,~\cite{corbett2016can,furqan2017learnability,srinivasan2019discovering}).
For example, if one issues a command to find two nodes, the system could suggest finding connections or finding the path as follow-up commands.}

\remove{Thus, an interesting research opportunity for future multimodal visualization systems (particularly those supporting speech) is to explore ways to help users discover both---which operations may help them address the task at hand and how they can perform desired operations.
To this end, one aspect for future systems to explore may be to suggest contextually-relevant operations based on the active state of the view and previously performed interactions.
For example, if one issues a command to find two nodes, the system could suggest finding connections and finding the path as follow-up commands.
In fact, depending on the presentation techniques, these suggestions could also be leveraged to help users discover \textit{how} they must perform operations.
For instance, when suggesting operations, the system could also display the touch gestures they can perform or command phrasings people could use~\cite{corbett2016can,furqan2017learnability,srinivasan2019discovering}.}

% \subsection{Enhancing feedback for speech commands}

% In the sessions that included speech input, participants often were confused about the active state of the visualization in response to a spoken command.
% Seven participants (P1-4, P7, P10, P12) even explicitly commented on this being a drawback of the system during their post-session interviews.
% For instance, P1 said that ``\textit{Because it [speech input] worked initially, I began to assume things worked. I didn't realize when it didn't actually do what I asked it to.}''
% Similarly, P10 said ``\textit{Couple of times I said something and it didn't select anything but I thought it did.}''
% Two participants (P2, P12) even assumed that it was their fault when the system did not respond correctly to a speech command.
% This raises an important point that visualization systems supporting natural language should clearly highlight and explain what changes have been made to the system in response to a query.
% Furthermore, systems should also give users the option to understand why a query failed and potentially even suggest related or alternative queries.

\subsection{Exploring Proactive Behavior and Supplementing Visualizations with Textual Summaries}

Recently, there has been a growing call for proactive system behavior in \edit{natural language interfaces (NLIs)} for visualization\remove{ NLIs}~\cite{srinivasan2017natural,tory2019mean}. 
% Srinivasan and Stasko~\cite{srinivasan2017natural} highlight value in `Exploring Proactive System Behavior' in NLIs for visualization.
One minor way in which Orko incorporates such behavior is by dynamically reordering the charts in the summary container (Figure~\ref{fig:system-ui}F) based on the user's most recent action~\cite{srinivasan2018orko}.
During the study, all six participants in the speech-only condition (P7-P12) and five other participants (P1, P2, P14, P17, P18) explicitly commented on this behavior being helpful.
Participants perceived the reordering of the summary charts as intelligent behavior and said that the charts often gave them answers for the questions they were thinking of posing next.
For instance, P18 said ``\textit{I really liked the charts that came up on the right. They always seemed to be relevant to what I was thinking of at the time.}"
Based on these observations, an open research opportunity lies in exploring more proactive multimodal visualization interfaces that preempt user questions.

Given the availability of speech as an input modalitiy, unsurprisingly, participants expected the system to be more conversational and even ``answer" questions.
For instance, advocating for support for textual responses in addition to changes in the visualization, P4 said ``\textit{Working with the system for a while starts making you want to ask higher-level questions and get specific answers or summaries as opposed to just the visualization}."
Based on such behavior, perhaps an interesting research opportunity is to explore multimodal network visualization systems that blend elements of question answering (QA) systems and also supplement visualizations with textual summaries.
% ~\cite{latif2018exploring,srinivasan2018augmenting}.
While recent work~\cite{latif2018vis,latif2019authoring} has begun exploring the idea of interactively linking text and network visualizations for presentation and storytelling,
extending these ideas to support interactive network exploration is an open area for future work.
\section{\edit{Limitations and Future Work}}

\textbf{Devices and Modalities.}~As common with laboratory experiments, our study had some limitations and constraints that must be considered when generalizing the results.
We only considered speech- and touch-based input in the context of a single vertical display located at a touching distance from a user.
Thus, building upon these results in different settings such as tablets or in AR/VR may require further testing.
Similarly, considering additional modalities such as pen or gaze may also have a major impact on participants' interactions and is another factor that we did not consider in the presented study.
We used Orko as our study interface since it was a minimalist system that supported core network visualization operations and was previously tested.
However, changing the system interface or interactions (e.g., including more sophisticated multi-touch gestures as in~\cite{schmidt2010set,thompson2018tangraphe}) may impact the interactions and participant preferences.

\vspace{.5em}
\noindent\textbf{Study Design and Datasets.}~As stated earlier, we did not counterbalance the order of the unimodal and the multimodal systems as participants interacting with the unimodal system after the multimodal system would already know and have experience with all the supported interactions.
That said, reversing the order of the systems could allow understanding what aspects of multimodal interaction participants ``missed" the most when working with the unimodal system.
While this is fundamentally different from understanding the effects of priming participants with one modality (\textbf{RG2}), it is certainly an important extension to the current study.
Furthermore, both the datasets used in the study were undirected, unipartite networks.
Although the operations covered in the study are generalizable, formally verifying the results and understanding potential variations in interaction patterns for dynamic, multipartite, and/or directed networks is an open topic for future work.

\vspace{.5em}
\noindent\textbf{Speech Recognition Errors.}~We used the standard speech recognition API~\cite{webspeechapi} for web-based systems and trained it with the potential keywords (e.g., `find', `filter', `path') and dataset-specific values to improve accuracy.
Even so, there were 115 speech-to-text errors (excluded from Table~\ref{tbl:interactions} to avoid double counting interactions) across the 18 participants.
\edit{Specifically, in the speech-only interface, across the six participants (P7-P12), there were 4-18 recognition errors (avg.~9), and in the multimodal interface, the number of errors ranged from 0-8 (avg.~3) across the 18 participants.}
These recognition errors led to some frustration among participants that may have impacted their interaction patterns.
% For example, P15 said ``\textit{Voice recognition worked fine for simple commands but for complex ones I had to switch to touch}."
For example, after encountering recognition errors in the speech-only interface, P9 switched to using more touch interactions in the multimodal interface, saying that ``\textit{The voice recognition would have to be improved a lot before I can feel comfortable using voice alone to control the system}."
While these are valid concerns, they are beyond the scope of our work and are imposed by the available technology.
That said, such issues make the results more practically applicable by mirroring interactions with general voice user interfaces where incorrect recognition is the most common type of error~\cite{myers2018patterns}.
In fact, these errors coupled with the earlier stated comments about complementing speech with touch further motivate the need to design multimodal systems that give users the ability to overcome errors of speech input or give them the freedom choose a different form of input.
\section{Conclusion}

We report a qualitative user study investigating how people interact with a network visualization tool using only touch, only speech, and a combination of the two.
In addition to verifying that participants prefer multimodal input over unimodal input for visual network exploration, we discuss the different factors driving these preferences such as freedom of expression, the complementary nature of speech and touch, and integrated interactions afforded by the combination of the two modalities.
We also report different interaction patterns participants employed to perform common network visualization operations, highlighting how people naturally adapt to new modalities that are more suited for an operation.
We hope the observations from this study can help designers of future systems better understand user interaction preferences, ultimately resulting in the creation of multimodal visualization systems that are more expressive than current tools and that support a fluid interaction experience.

\ifCLASSOPTIONcompsoc
  \section*{Acknowledgments}
\else
  \section*{Acknowledgment}
\fi

This work was supported in part by the National Science Foundation grant IIS-1717111.

\ifCLASSOPTIONcaptionsoff
  \newpage
\fi

\bibliographystyle{IEEEtran}
\bibliography{references.bib}

% \begin{IEEEbiography[{\includegraphics[width=1in,height=1.25in,clip,keepaspectratio]{mshell}}]{Michael Shell}
\begin{IEEEbiography}[\vspace{-2em}{\includegraphics[width=1in,height=1.25in,clip,keepaspectratio]{{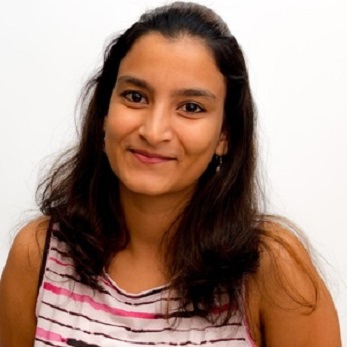}}}]{Ayshwarya Saktheeswaran}
is a Master's student in Human-Computer Interaction at the Georgia Institute of Technology. Her research interests include information visualization, human-computer interaction, and educational technology.
\end{IEEEbiography}

\begin{IEEEbiography}[\vspace{-2.5em}{\includegraphics[width=1in,height=1.25in,clip,keepaspectratio]{{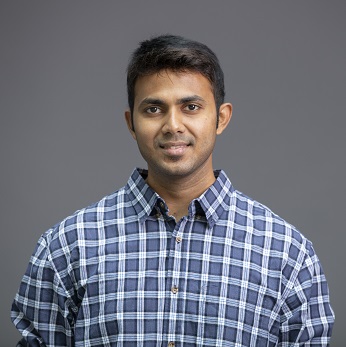}}}]{Arjun Srinivasan} is a Ph.D.~student in Computer Science at the Georgia Institute of Technology. His current research focuses on the design of intelligent and expressive visualization tools that combine multimodal input (e.g., speech and touch) and mixed-initiative interface techniques for human-data interaction.
\end{IEEEbiography}

\begin{IEEEbiography}[{\includegraphics[width=1in,height=1.25in,clip,keepaspectratio]{{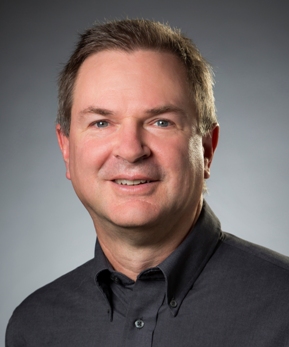}}}]{John Stasko}
is a Regents Professor in the School of Interactive Computing and the Director of the Information Interfaces Research Group at the Georgia Institute of Technology. His research is in the areas of information visualization and visual analytics, approaching each from a human-computer interaction perspective. John was named an ACM Distinguished Scientist in 2011 and an IEEE Fellow in 2014. He received his PhD in Computer Science at Brown University in 1989.
\end{IEEEbiography}

\end{document}